# Excitonic Spin-Coherence Lifetimes in CdSe Nanoplatelets Increase Significantly with Core/Shell Morphology

Phillip I. Martin,[†,1] Shobhana Panuganti,[†,1] Joshua C. Portner,[2] Nicolas E. Watkins,[1] Mercouri G. Kanatzidis,[1,3] Dmitri V. Talapin,[2,4] Richard D. Schaller*[1,4]

[1]Department of Chemistry, Northwestern University, Evanston, IL 60208, USA
[2]Department of Chemistry, University of Chicago, Chicago, IL 60637, USA
[3]Materials Science Division and [4]Center for Nanoscale Materials, Argonne National Laboratory, Lemont, IL 60439, USA

*Corresponding author: schaller@anl.gov, schaller@northwestern.edu

**Abstract**
We report spin-polarized transient absorption for colloidal CdSe nanoplatelets as functions of thickness (2 to 6 monolayer thickness) and core/shell motif. Using electro-optical modulation of co- and cross-polarization pump-probe combinations, we sensitively observe spin-polarized transitions. Core-only nanoplatelets exhibit few-picosecond spin lifetimes that weakly increase with layer thickness. Spectral content of differenced spin-polarized signals indicate biexciton binding energies that decrease with increasing thickness and smaller values than previously reported. Shell growth of CdS with controlled thicknesses, which partially delocalize the electron from the hole, significantly increases the spin lifetime to ~49 picoseconds at room temperature. Implementation of ZnS shells, which do not alter delocalization but do alter surface termination, increased spin lifetimes up to ~100 ps, bolstering the interpretation that surface termination heavily influences spin coherence, likely due to passivation of dangling bonds. Spin precession in magnetic fields both confirms long coherence lifetime at room temperature and yields excitonic g-factor.

**Introduction**

Colloidal II-VI semiconductor nanoplatelets (NPLs) constitute attractive candidates for a wide variety of optoelectronics owing to their unique band structure that derives from quantum confinement along a single direction in addition to narrow absorption and emission linewidths that arise from perfect ensemble thickness control with well-defined surface facets.[1-5] This material class, along with other morphologically similar semiconductors such as two-dimensional (2D)halide perovskites, have garnered interest for their unique spin polarization properties owing to highly controlled material features including tailored thickness of the inorganic semiconductor and, for the II-VI materials, established routes to replace synthetic organic ligands with dielectric inorganic shells.[6-10] In addition to tuning quantum well thickness, altering particle surface characteristics and controlling electron-hole overlap offer routes to influence spin-depolarization pathways, such as through interaction with surface dangling bonds, which may yield increased spin coherence lifetimes of benefit to spintronics.[11-13]

A recent study of colloidal 2D CdSe nanoplatelet quantum wells upon excitation with a circularly polarized pump and cross-polarized probe revealed a short-lived induced absorption at lower-energy wavelengths than the ensemble absorption feature.[2] The red-shifted absorption arises from Coulombic stabilization of biexcitons that show selectivity for the cross-polarized pump-probe combination over the period of time for which pump-induced spin polarization is preserved.

Relatedly, co-polarization of circular pump and probe leads to delayed growth of lower-energy differential induced absorption during the period in which scattering randomizes spin. Such spectral features enable characterization of lifetimes and mechanisms for spin-depolarization of photogenerated excitons that can help to develop understanding regarding fundamentals of exciton dynamics in these materials.[5, 9, 10, 14]

In this report, we perform circularly-polarized transient absorption spectroscopy on CdSe NPLs as well as CdSe/CdS and CdSe/ZnS core/shell NPLs in order to discern trends regarding the lifetime of spin-polarization upon increasing particle thickness or shell layer thickness. We observe that as NPL thickness increases from three to six monolayers (MLs) in core-only structures, the spin coherence lifetimes increase proportionally, but fitted lifetimes remain less than 2 ps; the 2ML CdSe NPL sample behaves anomalously, which may relate to appreciably altered lattice properties of this thinnest sample. Circularly polarized TA spectra for these structures convey biexciton binding energies when compared to static absorption, and here are found to be smaller than previously suggested. Spin-polarization lifetimes increase appreciably upon addition of CdS shell-layers to tens of picoseconds. CdS shell growth simultaneously yields surface modification in addition to reduced electron-hole overlap, both of which may increase spin lifetimes. Remarkably, we find that a single ZnS monolayer shell that does not appreciably alter electron-hole overlap also increases spin polarization lifetime significantly, suggesting the key importance of competing surface bonding effects that perhaps dominate electron-hole pair delocalization. Transient Faraday rotation measurements in static magnetic fields further confirm that observed signals arise from spin polarization, and precession frequencies afford exciton g-factor. These findings show that NPL morphology plays an appreciable role in stabilizing exciton spin orientation and can serve to decrease the availability of spin-decay pathways in ways that warrant further examination to achieve technological benefit.

**<u>Results and Discussion</u>**

Colloidal NPLs, synthesized according to previously published methods,[1-5, 15] were dispersed in hexane for presented optical measurements. Steady-state absorption spectra of the 2.5, 3.5, 4.5, 5.5, and 6.5 ML CdSe NPLs (herein referred to as 2CdSe, 3CdSe, 4CdSe, 5CdSe, 6CdSe, respectively) and CdSe/CdS core/shell NPLs with a 5CdSe core and 1, 3, or 6 ML CdS shell thickness (referred to as 5CdSe/1CdS, 5CdSe/3CdS, 5CdSe/6CdS, respectively) are shown in Figure 1. As the core-only NPL thickness increases, the lowest energy absorption transition redshifts (Figure 1a), while the CdSe/CdS core/shell particles further redshift from the core-only spectrum (for the same 5CdSe NPL core) with increasing shell-layer thickness as quantum confinement and dielectric confinement decrease (Figure 1b).

We performed transient absorption experiments using a 35 fs, amplified titanium:sapphire laser with a 2 kHz repetition rate. Pump pulses produced with an optical parametric amplifier were circularly-polarized using an achromatic zero-order waveplate, maintained at fluences corresponding to ~0.1 excitons per NPL on average, and made coincident on the sample together with mechanically time-delayed, circularly-polarized white light probe pulses. In order to minimize exciton spin-decay via hot carrier relaxation, samples were excited with pump wavelengths near respective absorption onset energies. Figure 2a shows the experimental configuration and representative transient absorption spectral maps versus pump-probe time-delay measured for the 5CdSe NPL sample. For cross-polarized data (Figure 2b), we observe a photo-induced absorption (PIA) feature near 553 nm that is initially intense and decays within several picoseconds. In the co-polarized TA spectral map (Figure 2c), the PIA is notably weaker initially but becomes stronger within the first few picoseconds. This low-energy PIA feature is consistent with previous reporting and confers spin-selective excitation of the singly excited, spin polarized exciton to a biexciton state.[8, 10]

To isolate spin-polarized features in the TA maps, transient data was collected using a mechanically chopped pump pulse that was co-polarized with the probe and then subtracted from separately measured cross-polarized data (Figure 2d). Manual subtraction of the independently collected datasets does successfully highlight spectral and temporal features related to spin polarized transitions but is also subject to appreciable noise in quantitative analyses of the exciton dynamics. To improve signal to noise, reduce data acquisition time, and implement a single multichannel detector for the probe pulses, we performed transient absorption measurements by replacing the typical mechanical chopper with an electro-optic inserted into the pump pulse train (see Figure 2a) set to produce half wave rotation at 1 kHz. Co- and cross-polarized pump-probe data were thus collected for immediately sequential pump pulses, not only eliminating the need to routinely collect probe-only spectra, but also decreasing background noise introduced from fluctuations in the pump or probe intensity over experimental lab time. Using this approach, we were able to directly generate polarization-differential transient spectral maps of the spin-dependent photophysical behaviors, as shown in Figure 2e.

Figure 3a shows the dynamic behavior of the spin-selective biexcitonic PIA across CdSe core-only samples as a function of increasing NPL thickness. These kinetic traces were then fit to single exponential decays to evaluate spin coherence lifetimes (Table S1). Over the range of increasing NPL thickness from 3CdSe to 6CdSe, spin-polarized decay lifetime increases from just 0.24 (±0.01) ps for 3CdSe up to 1.76 (±0.12) ps for 6CdSe, as shown in Figure 3b (data plotted to longer time-delays appears as Figure S1). A conceivable origin of this trend may be the reduced prominence of surface atoms as NPL thickness increases. Notably, the 2CdSe does not fit the otherwise systematic trend, instead exhibiting similar dynamic behavior as 5CdSe, suggesting that other contributing factors in the generation of long-lived spin-polarized species impact the particularly thin 2CdSe NPLs. This thickness dependence for core-only samples conveys lifetimes that are short and comparable to those recently reported for 2D perovskite spin-polarized lifetimes versus inorganic layer thickness.[7] Xiang *et. al.* reported fast, sub-picosecond decay components as evidence of hole spin-flips in CdSe with longer-lived spin polarization arising from particles that lack rapid hole trapping,[10] but relied on examining differences in dynamics for other exciton transitions that we did not observe. Notably, hole trapping has been implicated to occur on similar timescales as our observations[16] and could deleteriously affect excitonic spin polarization.

In addition to spin coherence decay lifetime, differenced spin-polarized transient *spectral* maps show clear changes between the lower-energy PIA in comparison to the heavy hole static absorption feature as a function of thickness, which relates biexciton binding energy (see Figure 3c). We find that attractive biexciton binding energy reaches ~50 meV in the 2CdSe core-only NPLs and decreases to ~15 meV in the 6CdSe core-only NPLs (Figure 3d). Reduction of biexciton binding energy is expected as particle thickness increases and trends in the direction of bulk-like behavior.[17] Such biexciton binding energies for 4CdSe have been estimated at up to 45 meV in the literature,[8] based in part on observation of biexciton-derived gain and lasing redshifts. However, we note that amplified spontaneous emission appears at energies prescribed by optical gain but is also influenced by optical loss from reabsorption. Measurements of binding energy reported in this report suffer less from the latter-described influence and relate a smaller value that is consistent both with other experimental data for 4CdSe[18, 19] and theory,[17] further highlighting the role of reabsorption from small Stokes shifts in gain and lasing efforts.

Figure 4 shows the spin-polarized dynamic behavior of CdSe/CdS core/shell NPL samples that were synthesized using the same 5CdSe core. Spin polarization lifetimes increase significantly for these core/shell structures. Decay kinetics are well-described by biexponential fitting with lifetimes $\tau_1$ and $\tau_2$ (see Fig. S2) for each sample exhibiting a rapid, ~1 ps process and a longer-lived, up to ~49 ps lifetime (see also Table S2). This longer decay component indicates a significant extension of spin-polarized exciton lifetime relative to the core-only samples. The $\tau_2$ lifetime increases sub-linearly with CdS shell thickness, which suggests diminishing influence on the preservation of spin polarization with both increased total volume of the nanostructure and increased average distance to the organic interface. Quantum confinement, it should be noted, also changes sub-linearly with shell layer thickness. Increases of the exciton spin coherence lifetime clearly arise from features of the core/shell morphology, but a challenge to mechanistic interpretation is that the core/shell system differs in multiple ways from the core-only particles. The core/shell samples experience reduced electron-hole wavefunction overlap for this combination of materials and reduced quantum and dielectric confinement in addition to altered CdSe interface chemistry. The quasi-type-II energy level alignment of the CdSe/CdS conduction bands permits electron delocalization into the CdS shell layer whereas holes nominally remain within the CdSe core.[13] Such delocalization could serve to reduce exchange interaction of the electron with the hole that can facilitate spin relaxation.[20] At the same time, isolated spins of dangling bonds and charges at the inorganic-organic interface may offer fast, efficient spin relaxation,[9, 21, 22] but the nearly-matched lattices at the inorganic-inorganic interface can largely reduce the importance of such relaxation processes.

To evaluate influences on CdSe/CdS core/shell spin lifetimes, we also prepared 5CdSe/ZnS core/shell NPLs with 1 or 3 ZnS monolayers. These structures, for which shells were carefully grown via colloidal atomic layer deposition using a fixed 5CdSe core,[5, 15] lack appreciable delocalization of either charge-carrier into this wider gap shell as noted by the absorption spectra shown in Figure 5a. ZnS shell growth does appreciably alter the surface bonding in comparison to organic ligand surface termination. Interestingly, the 5CdSe/1ZnS core/shell sample presented markedly slower spin depolarization relative to the core-only sample as shown in Figure 5b, with a significant amplitude ~100 ps lifetime as well as additional still longer-lived low-amplitude components (see Fig. S3 for biexponential fits and Table S3 for fit parameters). Upon continuing growth to a 3ZnS monolayer shell, lifetime decreases to ~21 ps perhaps influenced as strain builds up with the thicker, small-lattice-spacing shell. Overall, the significant lifetime increase suggests that removal of the proximal organic surface termination or passivation of dangling surface bonds for the core/shell structure serves to increase spin lifetimes in both the CdSe/CdS and CdSe/ZnS systems, consistent with theoretical investigations of Rodina *et al*.[21,22] Additionally, this finding suggests that the hole spin relaxation mechanism is not centrally important, as this charge remains contained within the CdSe core for the latter core/shell.

Finally, we performed transient Faraday rotation measurements in different magnetic fields[23] for the 5CdSe/1ZnS sample. Magnetic field causes the spin polarization to oscillate with the Larmor precession frequency associated with $E=g\mu_B B$. In these experiments field-induced oscillations occur and we find the oscillation frequency scales linearly with magnetic field strength with y-intercept near zero (~230 MHz). By fitting the period of oscillation versus field we obtain an excitonic g-factor of 1.83 that relates contributions of the electron and hole. This value is similar to other reports of exciton g-factor as studied using pump-orientation-probe measurements, such as in work by Feng *et. al*.,[24] and further confirms that the long lifetime observed indeed conveys spin coherence.

## Conclusion

Spin-polarization of excitons created with circularly-polarized laser pulses in core-only CdSe NPLs decay rapidly, occurring on picosecond timescales. Variation in sample morphology, such as particle thickness and surface passivation, are shown to impact and increase spin-depolarization lifetimes, with an increase from $\tau = 0.24$ ps in 3CdSe NPLs to $\tau = 1.76$ ps in 6CdSe NPLs. Growth of CdS shells yielded significant increases in spin-depolarization times, with an order of magnitude effect for thin-shelled samples of 5CdSe/1CdS NPLs ($\tau_2 = 24$ ps) up to $\tau_2 = 48.6$ ps for 5CdSe/6CdS NPLs. Additional measurements conducted on 5CdSe/ZnS samples, which do not alter electron-hole overlap would suggest a key role that surface passivation plays in maintaining excitonic spin coherence. In particular 5CdSe/1ZnS core/shell samples exhibited 100 ps lifetime, which is nearly two orders of magnitude longer than those observed in core-only samples. Studying these effects provides an important step towards understanding how to generate and maintain long-lived spin-polarized excitonic states in 2D semiconductors. Understanding specific spin-depolarization lifetimes and mechanisms will be valuable for the implementation of these states and materials in future technologies.

## Acknowledgements
Work performed at the Center for Nanoscale Materials, a U.S. Department of Energy Office of Science User Facility, was supported by the U.S. DOE, Office of Basic Energy Sciences, under Contract No. DE-AC02-06CH11357. We acknowledge student support provided by the National Science Foundation MSN under Grant No. 1808590. S.P. gratefully acknowledges financial support from the National Science Foundation Graduate Research Fellowship Program under Grant No. DGE-1842165 and the Ryan Fellowship. This work was supported in part by the National Science Foundation under Grant No. DMR-2019444 (IMOD an NSF-STC).

[†]These authors contributed equally to this work

## Figures

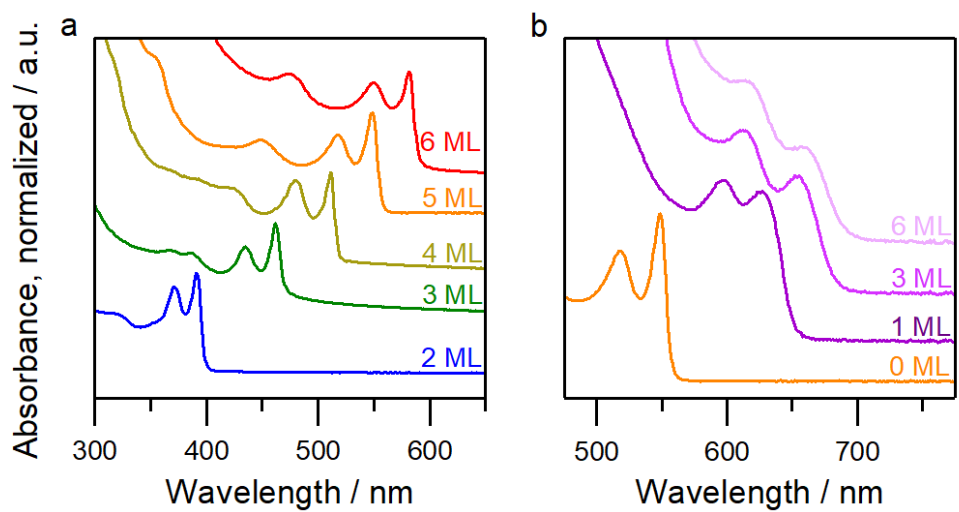

**Figure 1.** Absorption spectra of CdSe Nanoplatelet dispersions in hexane for **a)** indicated core-only monolayer thicknesses and for **b)** a 5CdSe NPL core onto which CdS shells of indicated layer thicknesses.

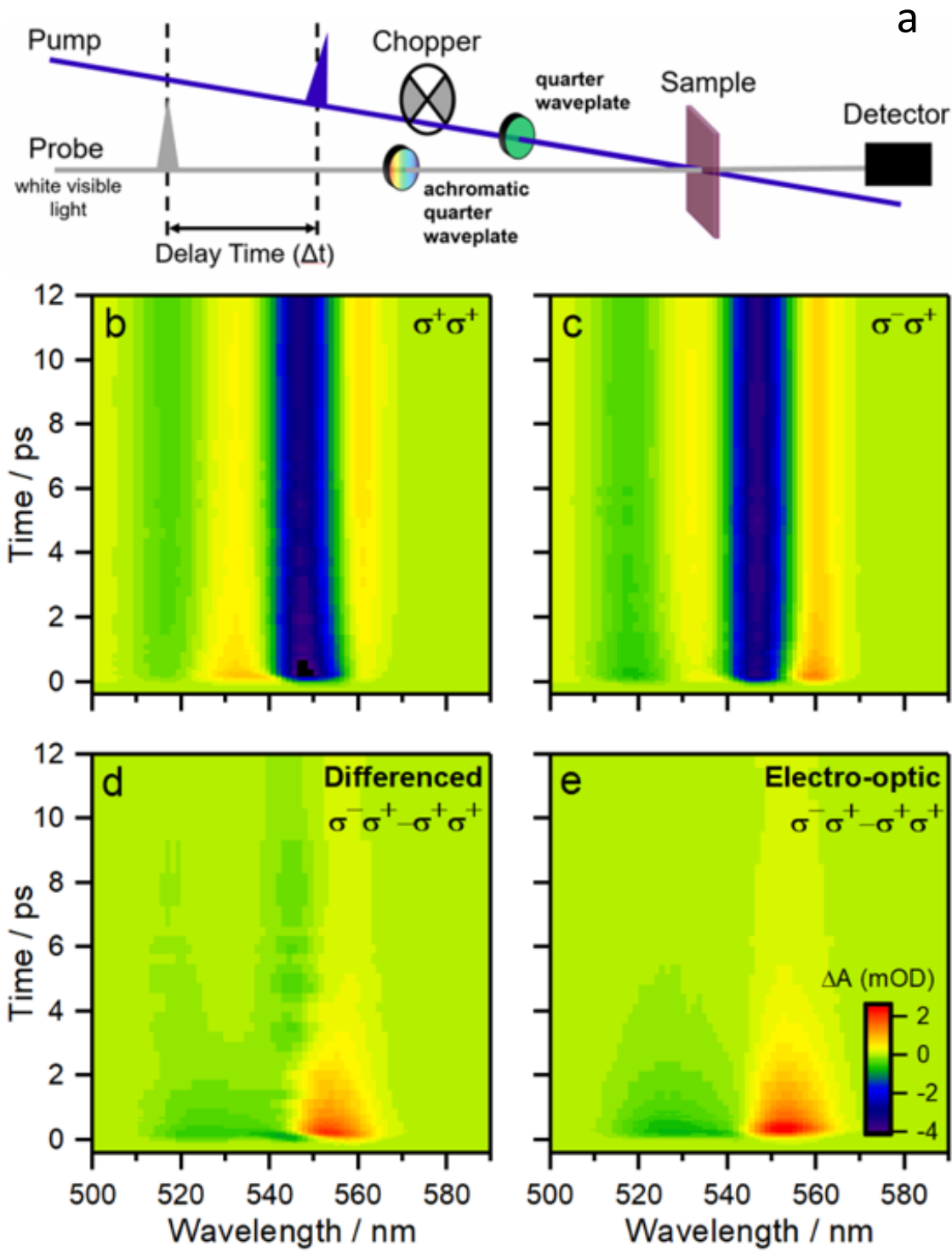

**Figure 2. a)** Schematic of circularly polarized transient absorption experiments. **b, c)** Transient absorption spectral maps for indicated pump and probe circular polarization conditions. **d)** Differencing panel b from panel c highlights time-dependent response of the polarization-

dependent spectral information. **e)** Replacement of the mechanical chopper with an electro-optical modulator directly yielded differential responses of the sample to co-polarized vs cross-polarized pump-probe conditions with increased signal to noise.

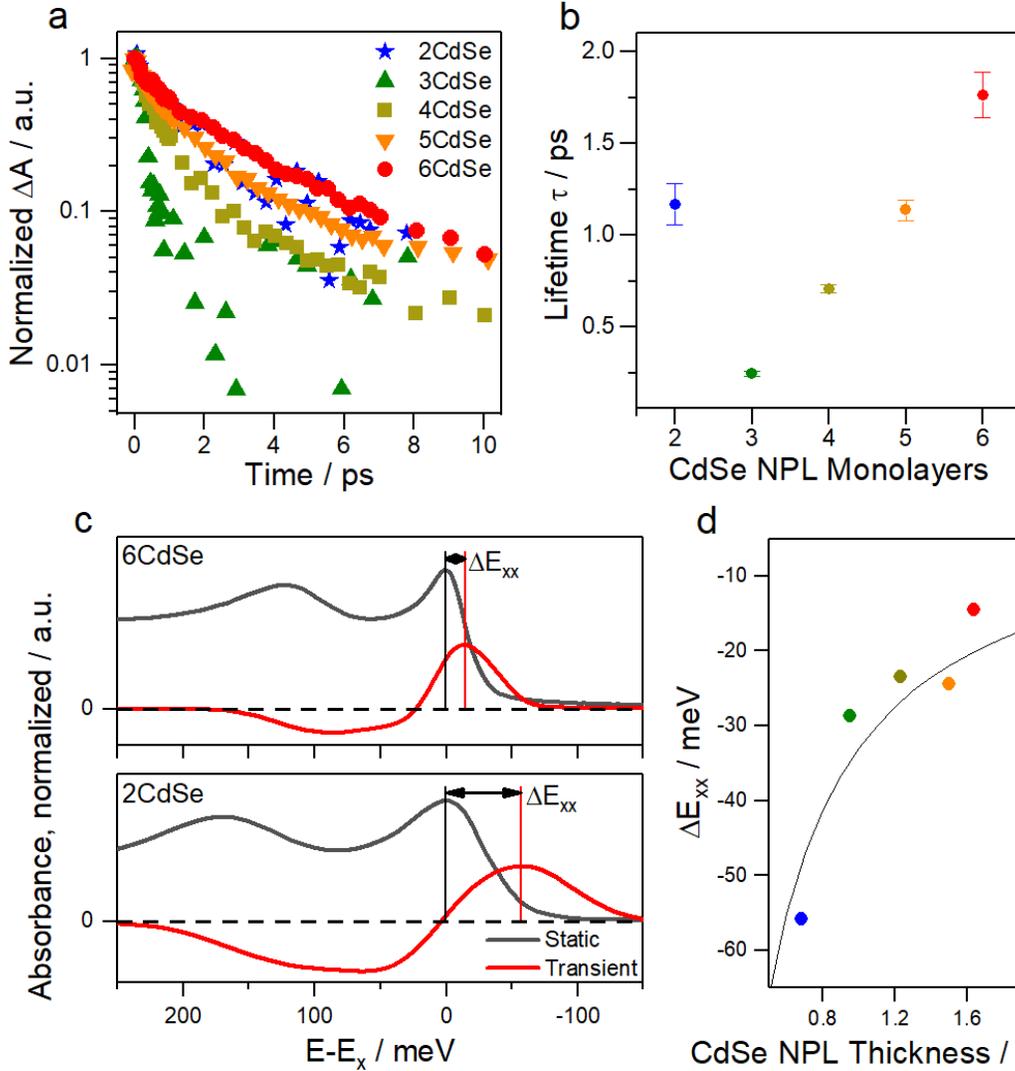

**Figure 3. a)** Spin polarization lifetime for indicated core-only NPL samples and **b)** fitted decay time constants. **c)** Biexciton binding energy, $\Delta E_{xx}$, evaluated from the spectral shift of the static absorption maximum to that of the transient spin polarized signal. **d)** Biexciton binding energy evaluated for indicated core-only NPL thicknesses spanning 2CdSe to 6CdSe.

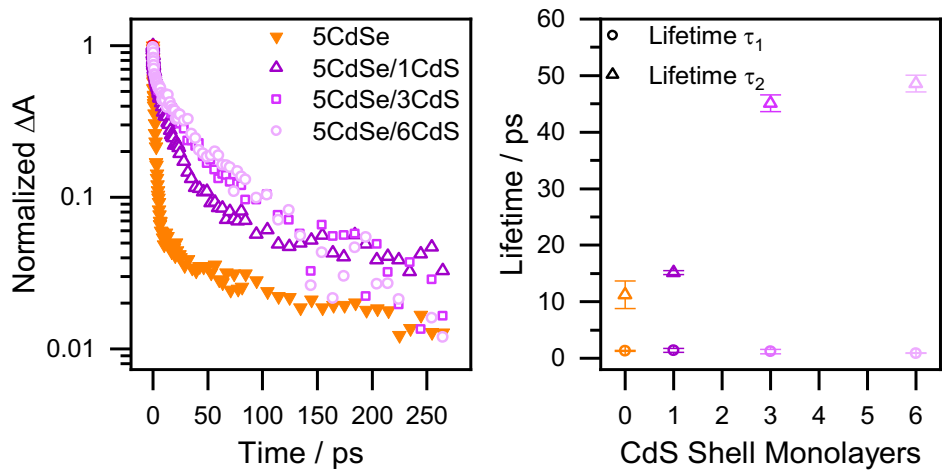

**Figure 4. a)** Dynamics of CdSe/CdS core/shell NPL samples that were synthesized using the same 5CdSe core. Spin polarization lifetimes increase significantly for these core/shell structures. **b)** Spin polarized decay kinetics are well-described by biexponential fitting with lifetimes $\tau_1$ and $\tau_2$.

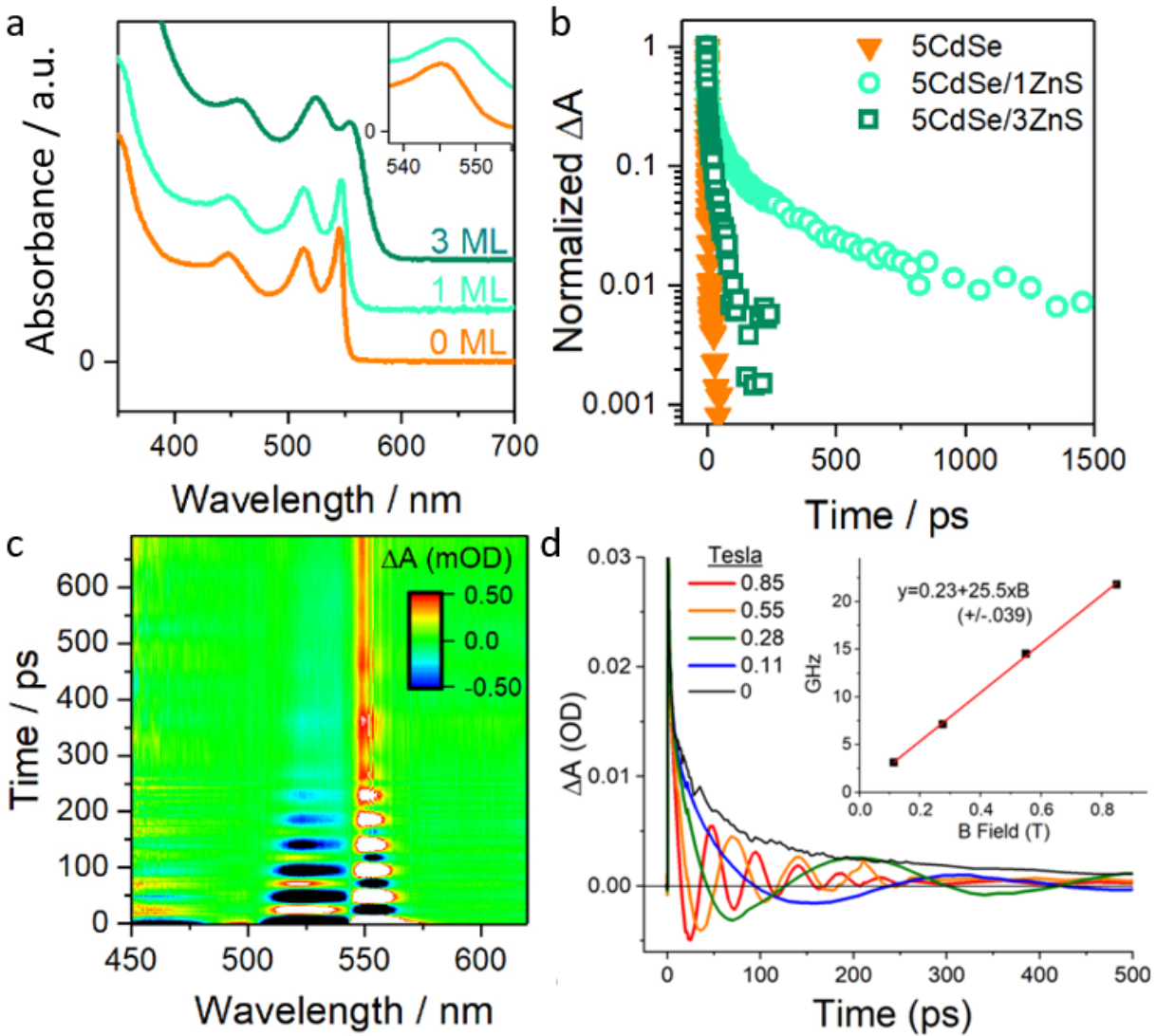

**Figure 5. a)** Absorption spectra of core/shell NPLs with a 5CdSe core and either 1 or 3 ZnS shell layers. **b)** Pump-polarization-modulated transient response of the three samples in panel A show an appreciable increase in spin polarized lifetime with even 1 layer of ZnS shell growth. **c)** Transient Faraday rotation of a 5CdSe/1ZnS nanoplatelet sample at room temperature shows oscillatory behavior with one discernible frequency. **d)** For the 5CdSe/1ZnS sample, measurement at several different indicated magnetic field strengths (here plotted at 554 nm probe wavelength) shows a linear dependence of oscillation frequency (inset), a g-factor of 1.83, and small y-axis intercept (230 MHz).

# Supplementary Information

## Excitonic Spin-Coherence Lifetimes in CdSe Nanoplatelets Increase Significantly with Core/Shell Morphology

Phillip I. Martin,[†1] Shobhana Panuganti,[†1] Joshua C. Portner,[2] Nicolas E. Watkins,[1] Mercouri G. Kanatzidis,[1,3] Dmitri V. Talapin,[2,4] Richard D. Schaller*[1,4]

[1]Department of Chemistry, Northwestern University, Evanston, IL 60208, USA
[2]Department of Chemistry, University of Chicago, Chicago, IL 60637, USA
[3]Materials Science Division and [4]Center for Nanoscale Materials, Argonne National Laboratory, Lemont, IL 60439, USA

*Corresponding author: schaller@anl.gov, schaller@northwestern.edu

Contents

**Materials and Methods**

**Chemicals.** 1-Octadecene (90%, Aldrich), Oleic Acid (90%, Aldrich), Oleylamine (technical grade, 70%, Aldrich), Selenium (100 mesh powder, 99.99%, Aldrich), Cadmium Nitrate Tetrahydrate ($Cd(NO_3)_2 \cdot 4H_2O$, 98%, Aldrich), Cadmium Formate ($Cd(HCOO)_2$, anhydrous, MP Biomedicals), Cadmium Acetate dihydrate ($Cd(OAc)_2 \cdot 2H_2O$, Aldrich), Lithium sulfide (98%, Strem), and Methylcyclohexane (anhydrous, Aldrich). Octadecene, oleic acid, and oleylamine were degassed at 100 °C and transferred to a nitrogen filled glovebox.

**Synthesis of 5 ML CdSe NPLs.** The synthesis of 5 ML CdSe NPLs was performed following previous literature.[1] Briefly, 170 mg of Cd(myristate)$_2$ in 14 mL octadecene (ODE) was degassed for 2 hours at 100 °C. Under nitrogen flow, the flask was heated to 240 °C and 1 mL of a 0.15 M Se solution sonicated in ODE was swiftly injected. After 30 seconds, 90 mg of Cd(acetate)$_2$ dihydrate was quickly added by lab spoon under a positive nitrogen flow and the flask was quickly resealed with a rubber septum. The solution was heated for 10 minutes with the temperature increased to 245 °C, then quickly cooled by air flow. At 100 °C, a solution of 2 mL oleic acid in 15 mL methylcyclohexane (MCH) was added and the solution was stirred under nitrogen overnight. The NPLs were isolated by centrifugation at 11000 rpm for 10 minutes. The resulting precipitate was redispersed in MCH and filtered through a 0.22 μm PTFE filter.

**Synthesis of 5CdSe/XCdS NPLs.** Cadmium sulfide shells were grown on core 5 ML NPLs by colloidal atomic layer deposition (c-ALD) based on previously reported methods.[2] Core NPLs (~20 mg) were precipitated from the stock solution in MCH with ethanol and transferring to a nitrogen filled glovebox. The NPLs were redispersed in 100 μL MCH and combined in a 4 mL vial with 500 μL ODE, 100 μL oleylamine, and 10 mg Li$_2$S with a Teflon coated stir bar. The vial was stirred at 150 °C on a hot plate (temperature measured by an ODE solution in an identical 4 mL vial) for 1 minute. After cooling, the Li$_2$S solids were removed by centrifugation and decantation, and additional NPLs were recovered from the precipitate with 50 μL MCH. The S$^{2-}$ capped NPL solution was then precipitated with 2 mL ethanol and redispersed in 100 μL MCH. The Cd layer was grown by adding the NPL solution to a 4 mL vial containing 500 μL ODE, 100 μL oleylamine, and 10 mg Cd(formate)$_2$ and stirring at 150 °C for 3 minutes. Any remaining solids were removed by centrifugation and decantation, then the solution was washed with 1 mL ethanol and redispersed in 100 μL MCH. This completed one monolayer of cadmium sulfide. Additional monolayers were grown by repeating the same procedure. Oleylamine and ODE were dried at 100°C for two hours under vacuum and transferred to a nitrogen filled glovebox. All steps were performed in a nitrogen glovebox.

**Synthesis of 5CdSe/XZnS NPLs.** Cadmium and zinc sulfide shells were grown on CdSe nanoplatelets using a modified c-ALD technique.[2] In the first half reaction, 10 mg anhydrous Li$_2$S, 50 µL oleylamine, 500 uL ODE, and ~10 mg CdSe NPLs in 100 µL MCH were combined and stirred at 150°C for 1 minute with a Teflon coated stir bar. The reaction was cooled to room temperature, and residual solids were removed by centrifugation. The NPLs in the liquid phase were purified by washing with ethanol and redispersing in MCH two times. In the second half reaction, 10 mg zinc acetate, 50 µL oleylamine, 500 µL ODE, and the S$^{2-}$ capped NPLs in 100 µL MCH were combined and stirred at 150°C for 5 minutes. The reaction was cooled to room temperature, and any residual solids were removed by centrifugation. The NPLs were purified by washing with ethanol one time and redispersing in 100 µL MCH. This completed one monolayer of zinc sulfide. Additional monolayers were grown by repeating the same procedure. Oleylamine and ODE were dried at 100°C for two hours under vacuum and transferred to a nitrogen filled glovebox. All steps were performed in a nitrogen glovebox.

**Spectroscopic Methods.** Static absorption spectra of the NPLs were collected in 2 mm quartz cuvettes using a Cary-50 spectrophotometer. Ultrafast experiments were performed using a Ti:sapphire amplifier with 2 kHz repetition rate and 35 fs pulse width. Pump pulses at were generated by frequency-dividing the 800 nm laser output using an optical parametric amplifier (OPA). A small portion of the 800 nm Ti:sapphire output was focused into a sapphire crystal to produce white light probe pulses. Probe pulses were mechanically time delayed using a translation stage and retroflector and circularly polarized using an achromatic quarter waveplate. Pump pulses were co- and cross-polarized relative to probe pulses using quarter waveplates and selected for every other pulse using a chopper. Additional measurements utilized a photoelastic modulator (electro-optic) operating at 2 kV to shorten collection time and improve data quality. Transient absorption spectra for each time point were averaged over 2 s, and three reproducible, separate scans were then averaged together. Background subtraction and probe chirp correction were conducted for all raw spectra before data analysis. Transient Faraday rotation measurements were performed using permanent magnetic fields that were applied perpendicular to the probe propagation direction. Field strengths at the sample position were evaluated with a Hall probe. All spectroscopic studies were performed at room temperature and in ambient conditions unless otherwise noted.

**Supplementary Data**

**Table S1.** Lifetimes of spin depolarization in 2, 3, 4, 5, 6 monolayer CdSe nanoplatelets from single-exponential fitting of spin decay.

| $ML$CdSe | $\tau$ (ps) | $\tau$ amplitude |
|---|---|---|
| 2CdSe | 1.17 ± 0.11 | 0.86 |
| 3CdSe | 0.24 ± 0.01 | 1.04 |
| 4CdSe | 0.71 ± 0.02 | 0.90 |
| 5CdSe | 1.14 ± 0.06 | 0.84 |
| 6CdSe | 1.76 ± 0.12 | 0.81 |

**Table S2.** Lifetimes of spin depolarization in 5CdSe/$X$CdS core/shell samples with $X$=1, 3, 6 monolayer CdS from biexponential fitting of spin decay.

| 5CdSe/$X$CdS | $\tau_1$ (ps) | $\tau_1$ amplitude | $\tau_2$ (ps) | $\tau_2$ amplitude |
|---|---|---|---|---|
| 5CdSe | 1.30 ± 0.05 | 0.93 | 11.24 ± 2.46 | 0.09 |
| 5CdSe/1CdS | 1.40 ± 0.21 | 0.35 | 24.0 ± 1.3 | 0.60 |
| 5CdSe/3CdS | 1.20 ± 0.06 | 0.39 | 45.1 ± 1.5 | 0.57 |
| 5CdSe/6CdS | 1.09 ± 0.09 | 0.39 | 48.6 ± 1.5 | 0.58 |

**Table S3.** Lifetimes of spin depolarization in 5CdSe/$X$ZnS core/shell samples with $X$=1, 3 monolayer ZnS from biexponential fitting of spin decay for larger amplitude components.

| 5CdSe/$X$ZnS | $\tau_1$ (ps) | $\tau_1$ amplitude | $\tau_2$ (ps) | $\tau_2$ amplitude |
|---|---|---|---|---|
| 5CdSe | 1.30 ± 0.05 | 0.93 | 11.24 ± 2.46 | 0.09 |
| 5CdSe/1ZnS | 2.5 ± 0.03 | 0.54 | 100.3 ± 3.8 | 0.31 |
| 5CdSe/3ZnS | 1.03 ± 0.05 | 0.96 | 21.4 ± 1.07 | 0.34 |

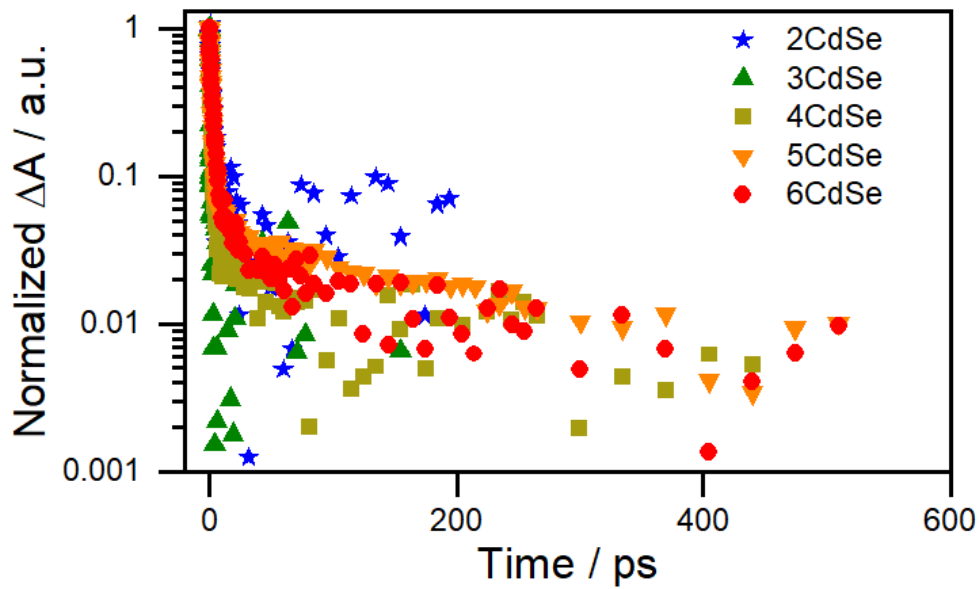

**Figure S1.** Dynamics of complete spin decay for indicated core-only CdSe samples ranging from 2CdSe to 6CdSe. The 2CdSe sample exhibits increased noise compared to other samples owing to the bluer probe wavelength.

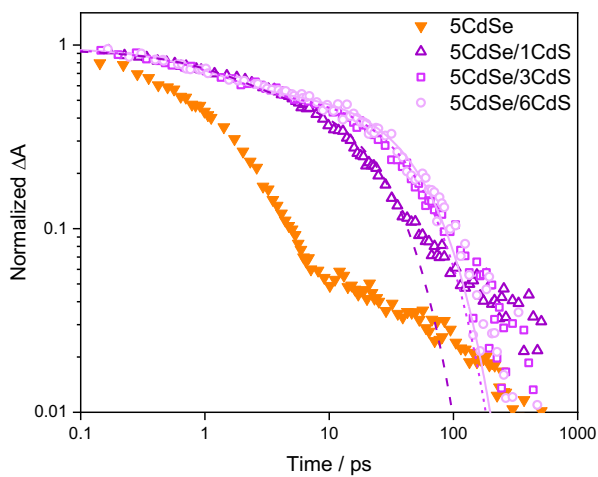

**Figure S2.** Spin depolarization for 5CdSe/$X$CdS samples. Biexponential fits, color coordinated with the data points are also show that are reflected in Table S2 above.

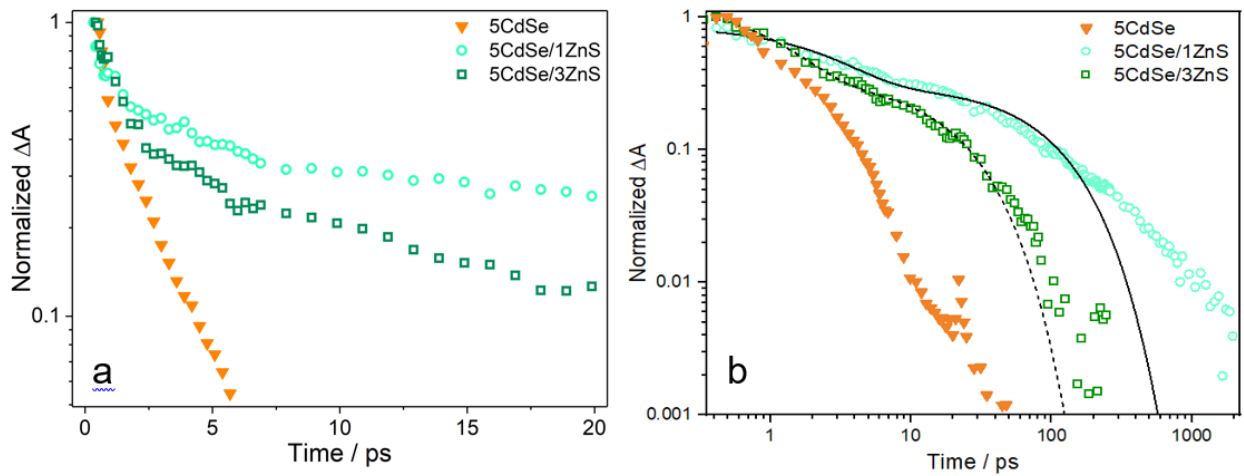

**Figure S3.** Spin depolarization for 5CdSe/*X*ZnS samples. **a)** Dynamics at early times show significant increase in lifetimes of spin-coherence in core/shell samples as compared to the core-only sample. **b)** Logarithmic scaling of kinetic traces highlights the significantly increased decay lifetime of the 5CdSe/1ZnS sample relative to both the 5CdSe core and 5CdSe/3ZnS sample by roughly an order of magnitude. Solid and dashed lines show biexponential fits with parameters that are presented in Table S3 above.